\documentclass{aa}
\usepackage{psfig}
\usepackage{graphics}
\def\lesssim{\mathrel{\hbox{\rlap{\hbox{\lower4pt\hbox{$\sim$}}}\hbox{$<$}}}}
\def\gtrsim{\mathrel{\hbox{\rlap{\hbox{\lower4pt\hbox{$\sim$}}}\hbox{$>$}}}}
\newcommand {\hh} {$\rm{h}^{-1} \,$ Mpc}
\newcommand {\h} {$\rm{h}^{-1} \,$ Mpc \,}
\newcommand {\ks} {km~s$^{-1} \;$}
\newcommand {\mpc} {Mpc$ \;$}

\newcommand {\gal}{$\xi _{\rm{gg}}(s) \;$}
\newcommand {\gr} {$\xi _{\rm{GG}}(s) \;$}

\newcommand {\galuw}{$\xi^{\rm{UW}} _{\rm{gg}}(s) \;$}
\newcommand {\gruw} {$\xi^{\rm{UW}} _{\rm{GG}}(s) \;$}
\newcommand {\ratm} {$<\xi_{\rm{GG}}/\xi_{\rm{gg}}> \;$}
\newcommand {\rat} {$\xi_{\rm{GG}}/\xi_{\rm{gg}} \;$}
\newcommand {\ratuwm} {$<\xi_{\rm{GG}}^{\rm{UW}}/\xi_{\rm{gg}}^{\rm{UW}}> \;$}
\newcommand {\ratuw} {$\xi_{\rm{GG}}^{\rm{UW}}/\xi_{\rm{gg}}^{\rm{UW}} \;$}

\begin{document}
\thesaurus{02(12.12.1; 11.03.1; 11.19.7)}
\title{The redshift--space two--point correlation function 
\\ of galaxy groups in the CfA2 and SSRS2 surveys}
\author{ Marisa Girardi\inst{1}
\and Walter Boschin\inst{2}
\and Luiz N. da Costa\inst{3,4}}
\offprints{M. Girardi (girardi@ts.astro.it)}
\institute{Dipartimento di Astronomia, Universit\`{a} degli Studi di Trieste,
Via Tiepolo 11, I-34100 - Trieste, Italy (girardi@ts.astro.it)
\and   Osservatorio Astronomico di Trieste, 
Via Tiepolo 11, I-34100 -
Trieste, Italy (boschin@ts.astro.it)
\and European Southern Observatory, Karl-Schwarzschild-Str.2, 
D-85748 Garching bei M\"unchen, Germany (ldacosta@eso.org)
\and Departamento de Astronomia CNPq/Observatorio Nacional, Rua General Jose
Cristino, 77, Rio de Janeiro, Brazil 20.921}
\date{Received 9 August 1999 /  accepted 5 November 1999}
\authorrunning{M. Girardi et al.} 
\titlerunning{The correlation function of galaxy groups}
\maketitle
\begin{abstract}
We measure  the two--point redshift--space correlation function of
loose groups of galaxies, \gr, for the combined CfA2 and SSRS2
surveys.  Our combined group catalog constitutes the largest
homogeneous sample available (885 groups).  We compare \gr with the
correlation function of galaxies, \gal, in the same volume.  We find
that groups are significantly more clustered than galaxies:
\ratm=$1.64\pm0.16$.  A similar result holds when we analyze a 
volume--limited 
sample (distance limit 78 \hh) of 139 groups.  For these
groups, with median 
velocity dispersion $\sigma_{\rm{v}}\sim 200$ \ks and mean group separation
$d\sim 16$ \hh, we find that the correlation length is 
$s_{\rm{0}}=8\pm1$ \hh,
which is significantly smaller than that found for rich clusters.  We
conclude that clustering properties of loose groups of galaxies are
intermediate between galaxies and rich clusters.  Moreover, we find
evidence that group clustering depends on physical properties of
groups: correlation strengthens for increasing $\sigma_{\rm{v}}$.
\end{abstract}
\section{Introduction} 

Loose groups of galaxies, the low--mass tail of the mass distribution
of galaxy systems, fill an important gap in the mass range from
galaxies to rich clusters.  Until now, clustering properties of loose
groups have been studied on the basis of rather small samples.  The
results are consequently uncertain and even contradictory.
Nevertheless, clustering properties of groups are shown to be robust
against the choice of the identification algorithm, provided systems
are identified with comparable number overdensity thresholds (Frederic
\cite{fredericb}).

The two--point correlation function, CF, of galaxies and of galaxy
systems constitutes an important measure of the large--scale
distribution of galaxies (e.g., Davis \& Peebles \cite{davis}; Bahcall
\& Soneira \cite{bahcall83}; de Lapparent et al. \cite{delapparent};
Tucker et al. \cite{tucker}; Croft et al. \cite{croft}).  From
galaxies to clusters the two--point correlation function in
redshift--space, $\xi (s)$, (e.g., Peebles \cite{peebles}), is
consistent, within errors, with a power--law form
$\xi(s)=(s/s_{\rm{0}})^{-\gamma}$ with $\gamma \sim 1.5-2$ for a variety of
systems.  The correlation length, $s_{\rm{0}}$, ranges from about 5--7.5 \h
for galaxies (e.g., Davis \& Peebles \cite{davis}; Loveday et al.
\cite{loveday}; Tucker et al. \cite{tucker}; Willmer et
al. \cite{willmer}; Guzzo et al. \cite{guzzo}) to $s_{\rm{0}} \gtrsim 15$ \h
for galaxy clusters (e.g., Bahcall \& Soneira \cite{bahcall83};
Postman et al. \cite{postman}; Peacock \& West \cite{peacock}; Croft
et al. \cite{croft}; Abadi et al. \cite{abadi}; Borgani et
al. \cite{borgani}; Miller et al. \cite{miller}; Moscardini et
al. \cite{moscardini}).

As far as loose groups are concerned, previous determinations of the
CF are very uncertain.  From the study of 137 groups (within CfA1) and
87 groups (within SSRS1), Jing \& Zhang (\cite{jing}) and Maia \& da Costa
(\cite{maia}) respectively find that the group--group CF, \gr, has a lower
amplitude than the galaxy--galaxy CF, \gal.  Analyzing 128 groups in a
sub--volume of CfA2N,  Ramella et al. (\cite{ramella90}; hereafter
RGH90) find that the amplitudes of \gr and \gal are consistent (see
also Kalinkov \& Kuneva \cite{kalinkov}).  Finally, Trasarti-Battistoni
et al. (\cite{trasarti}) study 192 groups in
the Perseus--Pisces region and find that the amplitude of \gr exceeds
that of \gal.

The theoretical expectations for the relative strength of the
group and galaxy  clustering are also contradictory.
Frederic (\cite{fredericb}) determines the correlation function for galaxy and
group halos in CDM numerical simulations by Gelb (\cite{gelb}) and finds that
groups are more strongly correlated than galaxies. In contrast,
Kashlinsky (\cite{kashlinsky}), on the basis of an analytical approach to the
clustering properties of collapsed systems of different masses,
concludes that groups and individual galaxies
should be correlated with the same amplitude.

Here we compute the two--point correlation function (in redshift
space) for 885 groups of galaxies identified in the combined CfA2 and
SSRS2 redshift surveys.  This sample is characterized by its large
extent (more than five times the volumes previously studied) and by
the homogeneity of the identification process (the
friends--of--friends algorithm FOFA; Ramella et al.
\cite{ramella97} -- hereafter RPG97).  Moreover, we compare the
group--group CF to that computed for galaxies in order to determine
the relative clustering properties of groups and galaxies. Because we
use the same galaxy sample where groups are identified, we avoid
possible effects of fluctuations due to the volume sampled.

In Sect.~2 we briefly describe the data; in Sect.~3 we describe the
estimation of the two--point correlation function; in Sect.~4 we compute
the correlation function of groups and compare it to that for
galaxies; in Sect.~5 we summarize our results and draw our conclusions.

Throughout the paper, errors are at the $68\%$ confidence level, and
the Hubble constant is 
$H_0=100\ \rm{h}\ $Mpc$^{-1}$ \ks.
\section{Galaxy and groups catalogs} 
 
We extract the sample of galaxies from the CfA2 North (CfA2N) and
South (CfA2S) (Geller \& Huchra \cite{geller}; Huchra et al.
\cite{huchra}; Falco et al. \cite{falco}), and the SSRS2 North 
(SSRS2N) and South (SSRS2S) (da Costa et al.  \cite{dacosta98})
redshift surveys.  These surveys are complete to $m_{\rm{B(0)}}\simeq 15.5$
and cover more than one--third of the sky, i.e. most of the
extragalactic sky.  The original papers contain detailed descriptions
of the observations and of the data reduction.  The velocities we use
are heliocentric; they include corrections for solar motions with
respect to the Local Group and for infall toward the center of the
Virgo cluster (see RPG97 for details). As in previous analyses of the
CfA2 surveys (e.g., Park et al. \cite{park}; Marzke et
al. \cite{marzke95}), we discard regions of large galactic extinction.
The total sample includes 13435 galaxies with radial velocity
$V<15000$ \ks.

 \begin{table}
      \caption[]{Data Samples}
         \label{tab1}

%
%

\begin{tabular}{lrr}
\hline \hline
\multicolumn{1}{c}{Sample}
&\multicolumn{1}{c}{$N_{\rm{G}}$}
&\multicolumn{1}{c}{$N_{\rm{g}}$}
\\
\hline
CfA2N &395& 5426\\
CfA2S &139& 2104\\
SSRS2N&123& 1697\\
SSRS2S&228& 3083\\
TOTAL &885&12290\\
\hline
\end{tabular}

   \end{table}
\begin{figure}
\psfig{figure=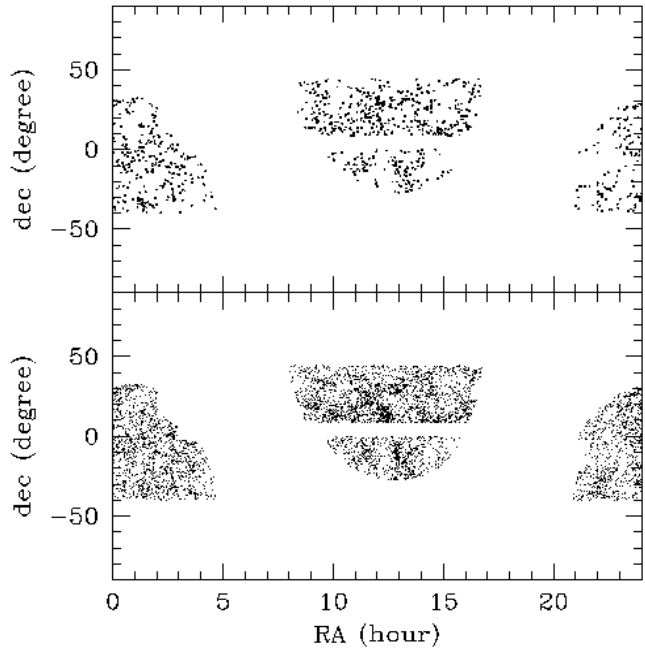,width=9cm,angle=0}
\caption{
We show the projection on the sky of the group and galaxy 
samples (upper and lower panel, respectively). 
}
\label{fig1}
\end{figure}

We use the catalogs of groups identified within CfA2N by RPG97 and
 within SSRS2 by Ramella et al. (in preparation).  The identification
 method is a friends--of--friends algorithm (FOFA) which selects
 systems of at least three members above a given number density
 threshold in redshift space.  In particular, RPG97 and Ramella et
 al. (in preparation) 
use the number density threshold $\delta\rho_{\rm{N}}/\rho_{\rm{N}}=80$
 and a line--of--sight link $V_{\rm{0}}=350$ \ks at the fiducial velocity
 $V_{\rm{f}}=1000$ \ks.  We run FOFA with these parameters on CfA2S and
 produce a group catalog for this survey, too.  The combined catalog
 contains a total of 885 groups that constitute a homogeneous set of
 systems objectively identified in redshift space.

The group catalogs are limited to radial velocities $V\le 12000$ \ks,
but members are allowed out to $V\le 15000$ \ks. We
confine the galaxy sample to $V\le 12000$ \ks
and are left with a total of 12290 galaxies.

Table~\ref{tab1} lists the numbers of groups, $N_{\rm{G}}$,
 and the numbers of
galaxies, $N_{\rm{g}}$, for each sample.  Fig.~\ref{fig1} shows the distribution of
the galaxy and group samples on the sky.

\section{Estimation of the correlation function} 

We compute the two--point correlation functions in redshift space for
groups and galaxies (hereafter \gr, and \gal, respectively).
The formalism in the two cases is the same.
We define the separation in the redshift space, $s$, as : 
\begin{equation} s=\frac{\sqrt{V_{\rm{i}}^2+V_{\rm{j}}^2-2V_{\rm{i}}V_{\rm{j}}cos\theta_{\rm{ij}}}}{H_{\rm{0}}},
\end{equation} 
\noindent 
where $V_{\rm{i}}$ and $V_{\rm{j}}$ are the velocities of two groups (or galaxies)
separated by an angle $\theta_{\rm{ij}}$ on the sky. Following
Hamilton (\cite{hamilton}) we estimate $\xi(s)$ with:
\begin{equation} \xi(s)=\frac{DD(s)RR(s)}{[DR(s)]^2}-1, \end{equation} 
\noindent where $DD(s)$, $RR(s)$, and $DR(s)$ are the number of
data--data, random--random, and data--random pairs, with separations in
the interval $(s,s+ds)$.
We build the control sample by filling the survey volume with a uniform 
random distribution of the same number of points as in the data. The points are
distributed in depth according to the selection function of the
surveys, $\Phi (V)$.

In order to decrease the statistical fluctuations in the determination of
$\xi(s)$, we average the results obtained using several different
realizations of the control sample. We compute 50
realizations in the case of groups and 5 in the case of galaxies.

Unless otherwise specified, we compute the ``weighted'' correlation
function by substituting the counts of pairs with $\sum w_{\rm{i}}w_{\rm{j}}$, the
weighted sum of pairs, which takes into account the selection effects
of the sample used.  In the case of a sample characterized
by the same selection function, $\Phi(V)$, volumes are
equally weighted and $w_{\rm{i}}=1/\Phi(V_{\rm{i}})$ is the weight of a 
group (or
galaxy) with velocity $V_{\rm{i}}$. The appropriate selection function
 for a magnitude--limited sample (e.g., de Lapparent et
al. \cite{delapparent}; Park et al. \cite{park}; Willmer et al. 
\cite{willmer}) is:
\begin{equation} \Phi(V)=\frac{\int^{M(V)}_{-\infty} \phi(M)
dM}{\int^{M_{\rm{max}}}_{\rm{-\infty}}\phi(M) dM}, \end{equation} 
\noindent where $\phi(M)$ is the Schechter (\cite{schechter}) form of the
luminosity function.  $M_{\rm{max}}$ is a low luminosity cut--off.  We chose
$M_{\rm{max}}=-14.5$, the absolute magnitude corresponding to the limiting
apparent magnitude of the survey at the fiducial velocity $V_{\rm{f}}=1000$
\ks. This value of $V_{\rm{f}}$ is the same as in RPG97.
The Schechter parameters of the galaxy luminosity
function, before the Malmquist bias correction,
are: $M^*=-19.1$, $\alpha=-1.1$ for CfA2, and
$M^*=-19.7$, $\alpha=-1.2$ for SSRS2 (Marzke et al. \cite{marzke94}; 
\cite{marzke98}).

\begin{figure}
\hspace{-0.5cm}\psfig{figure=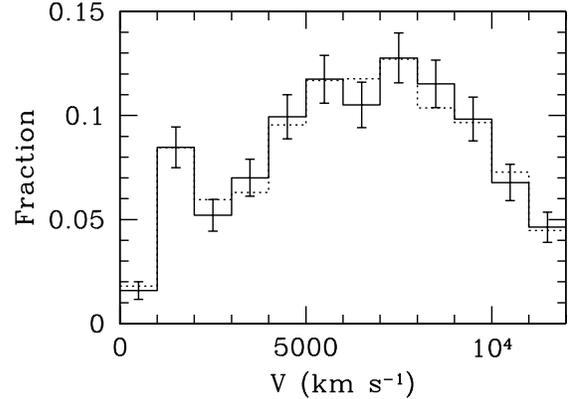,width=10.8cm,angle=0}
\vspace{-5.5cm}
\caption{
The velocity distributions of groups (solid line)  and of galaxies (dotted
line) are compared. 
The error bands represent $1-\sigma$ Poissonian errors. 
}
\label{fig2}
\end{figure}

We assume that the group selection function is the same as for
galaxies.  In fact, the velocity distributions of groups, $N_{\rm{G}}(V)$,
and of galaxies, $N_{\rm{g}}(V)$, are not significantly different according
to the Kolmogorov--Smirnov test (cf. Fig.~\ref{fig2}).  RGH90 and 
Trasarti-Battistoni et al. (\cite{trasarti}), and
Frederic (\cite{fredericb}) make the same assumption for observed and simulated
catalogs, respectively.

The different luminosity functions of CfA2 and SSRS2 correspond to
different selection functions. For  this reason we assign to a group
(galaxy) $i$, belonging to the subsample $k$, the weight given by:
\begin{equation}
w_{\rm{i}}=\frac{1}{\Phi_{\rm{k}}(V_{\rm{i}}) n_{\rm{k}}},
\end{equation}
\noindent where $n_{\rm{k}}$ is the mean number density of groups (galaxies) of that
subsample (e.g. Hermit et al. \cite{hermit}).  We compute the density
as $n_{\rm{k}}=1/{\cal V}_{\rm{k}} \cdot \Sigma _{\rm{i}} [1/\Phi (V_{\rm{i}})]$, where the sum is
over all the groups (galaxies) of the subsample volume, ${\cal V}_{\rm{k}}$ (Yahil et
al. \cite{yahil}). In our analysis, ${\cal V}_{\rm{k}}$ is the effective
volume of the subsample.  Because the different selection functions,
we also build a control sample for each subsample separately, and,
conservatively, we do not consider pairs of groups (galaxies) linking
two different subsamples.  In this way we also avoid crossing large
unsurveyed regions of the sky.

We compute the errors on $\xi(s)$ from 100 bootstrap re--samplings of
the data (e.g., Mo et al. \cite{mo}). Note that the
bootstrap--resampling technique, which overestimates the error in
individual bins, represents a conservative choice in this work.
\section{The group--group correlation function} 

\begin{figure}
\hspace{-0.5cm}\psfig{figure=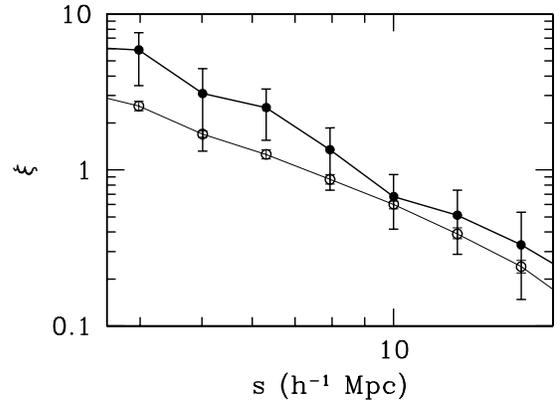,width=10.8cm,angle=0}
\vspace{-5.5cm}
\caption{
The group-group (filled circles) , and galaxy-galaxy (open circles) 
CFs, as computed for the combined 
sample. 
The error bands represent $1-\sigma$ bootstrap errors. 
}
\label{fig3}
\end{figure}

We plot the group--group CF, \gr, in Fig.~\ref{fig3}.  In the same figure we
also plot the galaxy--galaxy CF, \gal.  On small scales ($s\lesssim
3.5$ \mpc) \gr starts dropping because of the anti--correlation due to the
typical size of groups.  On large scales ($s \gtrsim 15$ \hh) the
signal--to--noise ratio of \gr drops drastically.  We thus limit our
analysis to the separation range $ 3.5 \lesssim s \lesssim 15$ \hh.

\begin{figure}
\hspace{-0.5cm}\psfig{figure=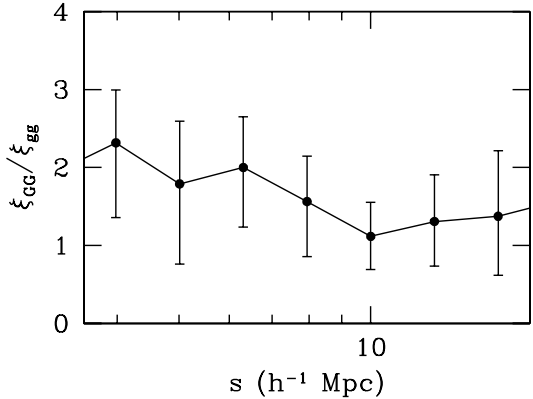,width=10.8cm,angle=0}
\vspace{-5.5cm}
\caption{
The ratio \rat, where the CFs 
are those computed for 
the combined sample in Fig.~3. 
}
\label{fig4}
\end{figure}

The main physical result in Fig.~\ref{fig3} is that \gr has a larger
amplitude than \gal.  This property of the CFs is also evident in
Fig.~\ref{fig4}, where we plot the ratio \gr / \gal on a linear scale.  Over
the $s$--range of interest, the values of the ratio are roughly
constant within the  errors. In order to give an estimate 
of the relative behavior of groups and galaxies we compute
the mean of the values of the ratio. We obtain \ratm=$1.64\pm0.16$.

\subsection{The CF of rich groups}

\begin{figure}
\hspace{-0.5cm}\psfig{figure=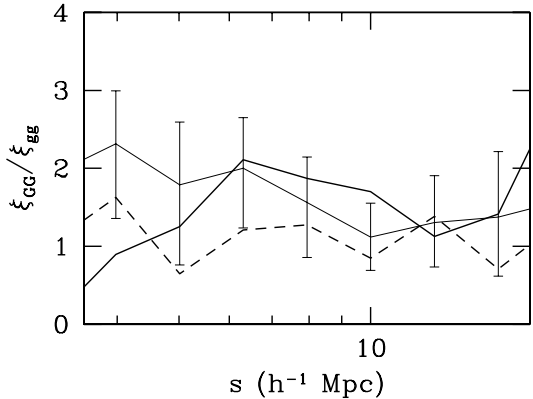,width=10.8cm,angle=0}
\vspace{-5.5cm}
\caption{
The ratio \rat for the combined sample 
(thin line),
but, now, \gr is computed also for groups with at least five member
galaxies (thick line) and for groups with less than five member
galaxies (dashed line).
}
\label{fig5}
\end{figure}

Groups with a number of members $N_{\rm{mem}}\ge 5$ are generally reliable,
as shown both by optical and X--ray analyses (Ramella et al. \cite{ramella95};
Mahdavi et al. \cite{mahdavi}).  On the other
hand, the reliability of groups with 
fewer members is often questionable.  In particular, the
analysis of the CfA2N survey performed by RPG97, and the analysis of a
CDM model by Frederic (\cite{frederica}) show that a significant 
fraction of the
triples and quadruples in group catalogs could be spurious.

We consider the 321 rich groups with $N_{\rm{mem}} \ge 5$, and find again
that groups are more correlated than galaxies.  Moreover, we find
evidence that the CF computed for richer groups is higher than the CF
computed for poorer groups, i.e.  $<\xi_{\rm{GG,rich}}/\xi_{\rm{gg}}>=1.48\pm0.16$
and $<\xi_{\rm{GG,poor}}/\xi_{\rm{gg}}>=1.10\pm0.14$ (cf. Fig.~\ref{fig5}).

The component of spurious groups among poor groups could be
responsible for the lower amplitude of the \gr of poor groups compared
to the \gr of rich groups. In fact, spurious groups should be
distributed like non--member galaxies.  However, at least part of the
observed higher clustering amplitude of rich groups could be due to
the existence of a clustering amplitude vs richness relationship.  The
relationship has been discussed for a variety of systems by several
authors (e.g. Bahcall \& West \cite{bahcall92}; Croft et al. 
\cite{croft}; Miller
et al. \cite{miller}).  Richness is
usually taken as a measure of the mass of the system, mass being the
real, interesting physical quantity directly related to the predictions
of cosmological models.

\subsection{The CF in the volume--limited sample}

\begin{figure}
\hspace{-0.5cm}\psfig{figure=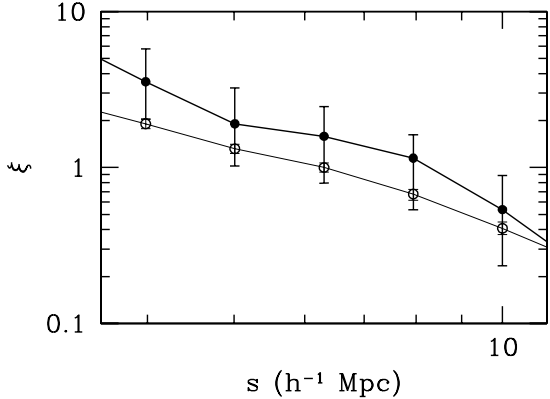,width=10.8cm,angle=0}
\vspace{-5.5cm}
\caption{
The group-group (filled circles) and galaxy-galaxy (open circles) 
CFs, as computed for the volume--limited sample.
}
\label{fig6}
\end{figure}

For our groups, richness is not a good physical parameter.  A better
parameter is the group (line--of--sight) velocity dispersion, $\sigma_{\rm{v}}$ (e.g. RPG97).
In a magnitude--limited sample any group selection based
on velocity dispersion will affect the selection function in an ``a
priori'' unknown way.  To avoid this problem, we analyze the 
volume--limited group sample built by Ramella et al. (in preparation)
who run an appropriately modified version of FOFA within 
volume--limited sub--samples of the CfA2 and SSRS2 galaxy surveys.  In
particular, we consider the 139 distance limited groups within $V\le
7800$ \ks, roughly corresponding to the effective depth of CfA2. We
cut the volume--limited galaxy catalogs in the same way. Within this
sample we compute the ``unweighted'' CF estimator, i.e. 
we set $w=1$ for all
groups/galaxies.  For this sample the useful $s$--range is
$3.5\lesssim s\lesssim 12$ \h (see Fig.~\ref{fig6}).

We find that the ratio \ratm of the total volume--limited sample
is \ratm$=1.58\pm0.10$, similar to that computed for the 
magnitude--limited sample (\ratm $\sim 1.6$).  This result reassures us about
the reliability of the selection function 
we assume for groups.

In order to check a possible dependence of \gr on $\sigma_{\rm{v}}$,
we divide the group volume--limited sample 
into two subsamples of equal size, one
subsample containing groups with $\sigma_{\rm{v}} \ge 214$ \ks, the other
including the remaining low velocity dispersion groups.  We find
that high--$\sigma_{\rm{v}}$ systems are more correlated than those
with low $\sigma_{\rm{v}}$ (\ratm=$2.14\pm0.37$ and \ratm=$1.29\pm0.17$,
respectively).  This evidence is in agreement with that found for
clusters, and suggests a continuum of clustering properties for all
galaxy systems.

In this context, it is appropriate to compare the groups of the 
volume--limited sample, characterized by the median 
velocity dispersion $\sigma_{\rm{v}}=214$ \ks and by the mean
group separation $d\sim 16$ \hh, to rich clusters ($\sigma_{\rm{v}}\sim
700$ \ks; $d\sim 50$ \hh; e.g. Zabludoff et al. \cite{zabludoff}; 
Peacock \& West \cite{peacock}).  We fit \gr to the form $\xi(s)=(s/s_{\rm{0}})^{-\gamma}$ with a
non--linear weighted least squares method and find $\gamma=1.9\pm0.7$
and $s_{\rm{0}}=8 \pm 1$ \hh. Note that groups show similar slope but
significantly smaller correlation length than 
optically or X--ray selected
clusters, for which
$s_{\rm{0}}\gtrsim 15$ \h (e.g. Bahcall \&
West \cite{bahcall92}; Croft et al. \cite{croft}; Abadi et al. 
\cite{abadi}; Borgani et
al. \cite{borgani}; Miller et al. \cite{miller}).  
Our results agree with the predictions of those N--body
cosmological simulations that also correctly predict the observed
cluster--cluster CF (e.g. cf. our ($s_{\rm{0}},d$) with Fig.~8
of Governato et al. \cite{governato}).

\subsection{The unweighted CF}

\begin{figure}
\hspace{-0.5cm}\psfig{figure=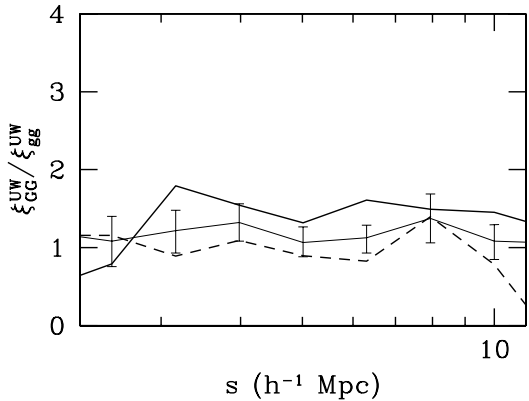,width=10.8cm,angle=0}
\vspace{-5.5cm}
\caption{
The ratio \ratuw for the combined sample (thin line),
and its nearby (dashed line) and distant (thick line) parts,
as computed with the unweighted CF.
}
\label{fig7}
\end{figure}
In order to verify the stability of our results against variations of
the weighting scheme, we compute the unweighted CF, $\xi^{\rm{UW}}$, for
the magnitude--limited sample.  We find that, as in the weighted
case, the amplitude of \gruw is still significantly higher than the
amplitude of \galuw, \ratuwm=$1.18\pm 0.05$.  We also find that the
amplitudes of \gruw and \galuw are both significantly lower than the
weighted estimates.

The differences between the results of the two weighting schemes rise
from the fact that the weighted CF weights each volume of space
equally and therefore better traces  the clustering of more distant
objects.  In fact, when we divide the group/galaxy catalogs in two
subsamples of equal size according to group/galaxy distances, the
distant samples ($V>6680$ \ks) give CFs with higher amplitude, i.e.
$<\xi^{\rm{UW}}_{\rm{GG,distant}}/\xi^{\rm{UW}}_{\rm{GG,nearby}}>=2.15\pm0.31$ and
$<\xi^{\rm{UW}}_{\rm{gg,distant}}/\xi^{\rm{UW}}_{\rm{gg,nearby}}>=1.43\pm0.08$.  Moreover,
the distant samples ($V>6680$ \ks) give
$<\xi^{\rm{UW}}_{\rm{GG,distant}}/\xi^{\rm{UW}}_{\rm{gg,distant}}>=1.43\pm0.12$ in closer
agreement with the result of the weighted analysis.  The ratio \ratuw
for the whole sample, as well as for its nearby and distant parts, is
shown in Fig.~\ref{fig7}.

As for a physical explanation, the fact that \galuw$<$\gal could be
the consequence of a dependency of clustering on luminosity, since the
unweighted CF estimator is more sensitive to the clustering of nearer,
fainter groups/galaxies (e.g., Park et al. \cite{park}).  In fact, the
dependency of clustering on luminosity has been pointed out for the
galaxy-galaxy CF (e.g., Benoist et al. \cite{benoist}; Cappi et
al. \cite{cappi}; Willmer et al. \cite{willmer}).  In addition, the
greater strength of \gal could be explained by different
clustering properties in different volumes of the Universe: e.g.,
Ramella et al. (\cite{ramella92}) find that the strength of 
the galaxy CF is very
high in the Great Wall. In the volume we examine the two biggest
structures, the Great Wall and the Southern Wall, both lie in distant
regions (e.g. da Costa et al.  \cite{dacosta94}) and therefore their
weight is larger in the weighted CF scheme.  It is reasonable to
expect also that  distant groups, which are brighter (and presumably
more massive) and which preferably lie in the two big structures, are more
strongly correlated than nearby groups leading to the observed
\gruw$<$\gr.
\section{Summary and conclusions}

We measure the two--point redshift--space correlation function of
loose groups, \gr, for the combined CfA2 and SSRS2 surveys.  Our combined
group catalog constitutes the largest homogeneous sample available
(885 groups).  We compare \gr with the correlation functions of
galaxies, \gal, in the same volumes.

Our main results are the following:
\begin{enumerate}
\item Using the whole sample we find that groups are significantly more
clustered than galaxies, \ratm=$1.64\pm0.16$, thus consistent
with the result by Trasarti-Battistoni et al. (1997), based on a much
smaller sample. This
ratio  can be considered a lower
limit considering the possible presence of unphysical groups.
\item Groups are significantly less clustered than clusters. In
particular, we find $\gamma=1.9\pm0.7$ and $s_{\rm{0}}=8\pm1$ for 139 groups
identified in a volume--limited sample ($V\le 7800$ \ks, median
velocity dispersion $\sigma_{\rm{v}}\sim 200$ \ks, and  mean group
separation $d\sim 16$ \hh). This result can be compared with that of
galaxy clusters ($s_{\rm{0}}\sim 15$--$20$ \h for systems with $\sigma_{\rm{v}}\sim 700$
 \ks and $d\sim 50$ \hh; e.g., Bahcall \& West \cite{bahcall92}; 
Croft et al. \cite{croft};
Abadi et al. \cite{abadi}; Borgani et al. \cite{borgani}; Miller et al. 
\cite{miller}).
\item There is a tendency of clustering amplitude to  increase with
group velocity dispersion $\sigma_{\rm{v}}$, which is the better indicator
of group mass at our disposal.  
\end{enumerate}
We conclude that there is a continuum of clustering properties of galaxy
systems, from poor groups to very rich clusters, with correlation
length increasing with increasing mass of the system.

\begin{acknowledgements}

We thank Stefano Borgani, Antonaldo Diaferio,
and Margaret Geller for useful discussions.
Special thanks to Massimo Ramella for enlightening suggestions.
M.G. wishes to acknowledge Osservatorio Astronomico di Trieste
for a grant received during the preparation of this work.
This work has been partially supported by the Italian Ministry of
University, Scientific Technological Research (MURST), by the 
Italian Space Agency (ASI), and by the Italian Research Council
(CNR-GNA).

\end{acknowledgements}

{}

\end{document}